\documentstyle[prl,aps,graphics,twocolumn]{revtex}

\begin{document}


\title{Event-by-Event Analysis of Proton-Induced Nuclear 
Multifragmentation: Determination of Phase Transition Universality-Class 
in System with Extreme Finite-Size Constraints}

\draft

\author{M. Kleine Berkenbusch,
  W. Bauer\cite{f1}, K. Dillman, S. Pratt}
\address{National Superconducting Cyclotron Laboratory and Dept.\
  of Physics and Astronomy, Michigan State
  University, East Lansing, Michigan 48824-1116, USA}
\author{L. Beaulieu\cite{f2}, K. Kwiatkowski\cite{f3}, 
  T. Lefort\cite{f4}, W.-c. Hsi\cite{f5}, V.E. Viola}
\address{Dept.\ of Chemistry and IUCF, Indiana University,
  Bloomington, IN 47405, USA}
\author{S. J. Yennello}
\address{Dept.\ of Chemistry and Cyclotron Laboratory, Texas A\&M
  University, College Station, TX 77843, USA}
\author{R. G. Korteling}
\address{Dept.\ of Chemistry, Simon Fraser University, Burnaby,
  BC, Canada, V5A 1S6}
\author{H. Breuer}
\address{Dept. of Physics, University of Maryland,
  College Park, MD 20742}

\date{\today}
		
\maketitle

\begin{abstract}
A percolation model of nuclear fragmentation is used to interpret 10.2
GeV/c p+${}^{197}$Au multi-fragmentation data.  Emphasis is put on
finding signatures of a continuous nuclear matter phase transition in
finite nuclear systems.  Based on model calculations, corrections
accounting for physical constraints of the fragment detection and
sequential decay processes are derived.  Strong circumstantial
evidence for a continuous phase transition is found and the values of
two critical exponents, $\sigma = 0.5 \pm 0.1$ and $\tau = 2.35 \pm
0.05$, are extracted from the data.  A critical temperature of $T_c =
8.3 \pm 0.2$ MeV is found.
\end{abstract}



Reactions in which excited nuclei break up into intermediate size
fragments, nuclear multi-fragmentation reactions, are believed to be
associated with a liquid-gas type phase transition in nuclear
matter.  However, so far no unambiguous proof for this transition has  
been found in experimental data. This is due to, primarily, the 
exteme finite size effect involved in systems of only on the order of 
$10^2$ constituents, the impossibility to fix a system at points in 
the phase diagram and study it there, as well as the complication due 
to sequential decays of the fragments produced in their excited state.

In this Letter we report results of an analysis of data on
proton-induced fragmentation reactions of a $^{197}$Au target at
incident energies of 10.2 GeV/c.  These data were collected by the
ISiS collaboration \cite{Lef99} in experiments at the Brookhaven
National Laboratory AGS accelerator facility.  A comparison with
percolation-theory-based models is conducted.  This comparison enables
us to pay particular attention to detector efficiency effects, finite
size effects, as well as to the role played by sequential decay
processes.  With these corrections applied, an event-by-event scaling
analysis is performed in order to derive values for the critical
exponents $\sigma$ and $\tau$ and the critical temperature $T_c$ of
the phase transition.

The ISiS collaboration has produced one of the most complete experimental
multifragmentation data sets with very high statistics. These data 
have also been interpreted in the framework of other phase transition 
models, in particular the SMM, EES, and SIMON models \cite{models},
usually associated with a liquid-gas type phase transition. However, within
the models the order of the phase transition depends on certain model
parameters, as well as on the size of the system, see for 
example \cite{Hau00}.  The percolation approach provides an effective
alternative way for determining the order of the phase transition and 
the influence of finite size effects.

The percolation model of nuclear multi-fragmentation used in our
analysis has been introduced originally by Bauer et al.  \cite{Bau84}
and used by many groups \cite{others}.  It utilitizes a representation
of the target nucleus by sites of a simple cubic lattice
($\mathbf{Z}^3$) in an approximately spherical shape, with
nearest-neighbor bonds representing the (short-ranged) strong force
between the nucleons.  These bonds are broken statistically
independently with a probability $p_b$.  Clusters of connected sites
are counted and interpreted as fragments.

Multi-fragmentation reactions, in particular proton-induced
multi-fragmentation reactions, can be thought of as three-step
processes. In the first step excitation energy is deposited in the
target nucleus and pre-equilibrium particles are emitted. In the
second step, the thermalized source breaks up into intermediate mass
fragments. In the final
step, the excited pre-fragments decay via standard sequential decay
channels into the fragments that can be observed by the detector. 

Step One: Energy deposition.  The percolation model needs a bond
breaking probability $p_b$ as input.  $p_b$ can be determined from the
energy deposited in the system via
\begin{equation}
  p_b(E^*) = 1- \frac{2}{\sqrt{\pi}} \Gamma \left( \frac{3}{2}, 0,
    \frac{B}{T(E^*)} \right)
\end{equation}
where $\Gamma(x, z_0, z_1)$ is the generalized incomplete Gamma
function, $B$ is the binding energy per nucleon in the source, $T$ is
the temperature of the source, and $E^*$ is the excitation energy per
nucleon of the source \cite{Li9394}.  It is assumed that the relation
between the excitation energy $E^*$ of the fragmenting source and the
temperature is given by $E^*=a T^2$ with $a=A_0/13$ (corresponding to
the high temperature limit of a degenerate Fermi gas model; $A_0$ is
the mass number of the residue nucleus, compare also \cite{Ell96}). 
Here we utilitize the energy deposition and source size information as
determined from the experimental data \cite{Lef99}.
It can be argued that $a=A_0/8$ should be used for low excitation 
energies, where surface effects are dominant.  However, close to the 
critical point surface effect disappear, and this motivates our choice 
of $a=A_0/13$.  One should keep in mind, however, that this choice 
will have some (minor) consequences for the exact value of the 
critical value of the control parameter of the percolation model.

Step Two: Percolation.  For the size of the lattice, we use the charge
of the nucleus after emission of pre-equilibrium particles.  This
information is also contained in the ISiS data set, on an 
event-by-event basis.  Alternatively,
one could also use the mass of the source.  For the theory, this
provides no difficulties whatsoever.  However, in the experiment mostly
the charges of the particles are detected.  Thus it is natural to use
the charge as the relevant quantity in our calculations.  The
assumption that the source has an approximately spherical shape after
the emission of pre-equilibrium particles is supported by nuclear
transport theory calculations in the BUU model \cite{WKV96}.
After setting up the source on the lattice, the bonds are broken with
the probability $p_b$ and the cluster structure is analyzed.

Step Three: Sequential Decay.  For this,
we use a computer code we recently developed to investigate radioactive
isotope yields in RIB facilities \cite{PBM01}.  Eight decay modes were
considered: proton, neutron, deuteron, dineutron, diproton, t,
${}^3$He and $\alpha$.  The decay weights were chosen according to
Weisskopf arguments.  For nuclei up to nitrogen experimentally
measured values were used for the excited states.  Decays were
calculated for all levels in all nuclei.  For the decay of each level,
the decay rate was calculated into every possible level energetically
accessible through the decay modes listed previously.  The weight
associated with the decaying nucleus was then apportioned into all the
states in proportion to the rates for the decay into such states.  The
weights were also simultaneously added into the ground states of the
nuclei representing the decay modes.  Thus, the decaying process
exactly preserved the initial $N$ and $Z$ of the original source system.

While the ISiS data set contains essentially complete events, it is
still subject to the usual problems associated with multi-particle
detector systems of subatomic particles, such as energy cuts, gaps
between the active areas of the detector elements, loss of charge and
mass-resolution for heavier fragments, and fragments that escape
detecting by being stopped in the target or traveling down
the beam pipe.  For the quantitative study we attempt here, these
effects can not be neglected.  We have thus created extensive filter
software to simulate detector acceptance effect.

\begin{figure}
 \resizebox*{0.8\columnwidth}{!}{\includegraphics{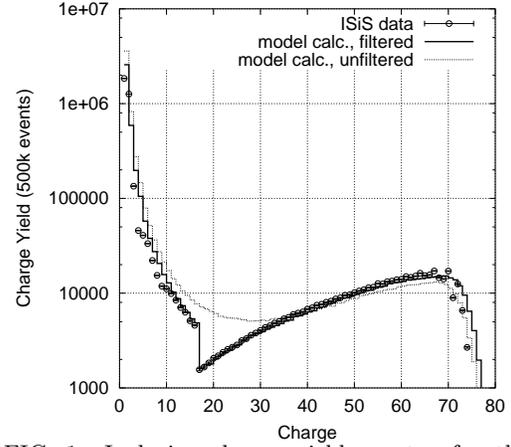}}
  \caption{Inclusive charge yield spectra for the reaction $p+Au$ at
    10.2 GeV. The round plot symbols represent the ISiS dat. The
    dotted histogram is the result of the corresponding percolation
    model calculation. The thick histogram represents the output of
    the calculation, filtered through the detector acceptance 
    corrections.}
  \label{fig:chargeyield}
\end{figure}

Fig.  \ref{fig:chargeyield} shows the comparison of our calculations
with the experimental data.  The data points with the (very small)
error bars represent the results of the experimental charge yields. 
The discontinuity at charge $Z=17$ is due to the fact that only
charges up to that value could be resolved elementally by the detector
and the assumption that all missing mass is contained in a single
residue (corrected for prompt particle emission during the fast
cascade stage of the reaction).  The dotted histogram is the result of
our model calculations, as described in the previous section, without
applying the filter.  Filtering of our model calculations through the
detector acceptance filter yields the thick histogram.  It is in
essentially perfect agreement with the data.  The discrepancy between
the two histograms thus gives us a good understanding of the degree to
which the raw experimental data are affected by detector
acceptance effects.
One can also investigate more exclusive observables, such as the 
vanishing of the largest cluster as a function of the multiplicity, or 
the second moment as a function of the multiplicity.  For the 
percolation model, these comparisons were published previously for 
other, but similar, data sets \cite{Bot,Fried}.  Here, we obtain 
similar degree of agreement.  These comparisons, as well as a 
comparison of a charge of the largest cluster for different 
multiplicity bins, have been performed \cite{thesis} and will be 
published in a forthcoming paper.

From analytical solutions and numerical results on very large
lattices, it can be inferred that in percolation theory, for the
control parameter $p$ assuming values close to the critical value
$p_c$, the cluster numbers scale as
\begin{equation}
  n_s(p) = s^{-\tau} f \left[ (p-p_c) s^\sigma \right] \qquad
  (\textrm{for } p \approx p_c)
\label{equ:clusterscaling}
\end{equation}
where $s$ is the size of a cluster.
The scaling function $f$ has the property $f(0)=1$ and accounts for
the fact that a power law dependence is only correct in the case of
$p=p_c$. 

Implicitly introduced by Eq.~\ref{equ:clusterscaling} are two
critical exponents of percolation theory: $\sigma$ and $\tau$. With
the definition $s_\xi = (p-p_c)^{-1/ \sigma}$, we can rewrite
Eq.~\ref{equ:clusterscaling} as:
\begin{equation}
  n_s(p) = s^{-\tau} f \left[ \left( \frac{s}{s_\xi} \right)^\sigma
  \right]
\label{equ:clustersizescaling}
\end{equation}
This leads to the interpretation of $s_\xi$ as a crossover size for
the cluster sizes from power law abundance for $s \ll s_\xi$ to
exponentially rare clusters of size $s \gg s_\xi$. 

\begin{figure}
 \resizebox*{0.8\columnwidth}{!}{\includegraphics{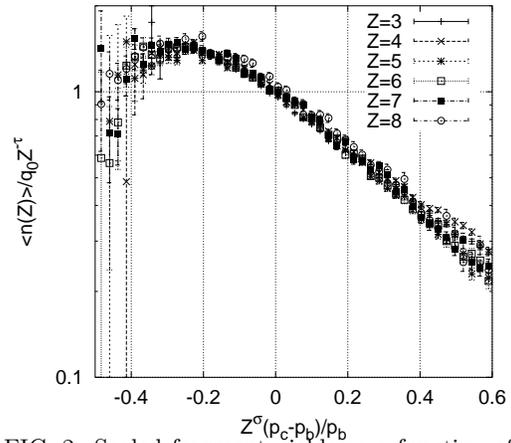}}
 \caption{Scaled fragment yields as a function of the scaled control
   parameter for the model calculations. The yields for
   the $Z$ = 3, 4, 5, 6, 7, 8 fragments are shown.}
 \label{fig:scaled_perc}
\end{figure}

A special case of the general Eq.~\ref{equ:clusterscaling} is,
for example, the scaling implied by the Fisher droplet model
\cite{Fis67},
\begin{equation}
  \langle n_Z \rangle = \left\langle \frac{N_Z}{Z_0} \right\rangle = q_0
  Z^{-\tau} \exp{ \left[ \frac{Z \Delta \mu}{T} - \frac{c_0 \epsilon
        Z^\sigma}{T} \right] } 
\label{equ:fdmscaling}
\end{equation}
where $Z$ is the size of a droplet and $\epsilon=(T_c-T)/T$ is the
scaled control parameter.  From this equation we expect a straight
line when plotting $\langle n_Z \rangle / q_0 Z^{-\tau}$ in a semi-log 
plot (compare Fig.~~\ref{fig:scaled_perc}) vs.\
$\epsilon Z^\sigma$ in the vicinity of the critical point, provided
the scaling behavior holds.  In addition, the straight line should
have the property of $f(0)=1$.  (Here it is assumed that the bulk
factor $\exp{ [Z \Delta \mu / T] }$ is close to unity, an assumption
that is supported by earlier findings of Elliott et al.~\cite{EMP00}.)

In the context of the percolation model, the same scaling behavior can
be expected if one substitutes the temperature $T$ by the bond
breaking probablity $p_b$.  Again, the cutoff function $f$ in the
scaling equation is then given by the exponential factor in
Eq.~\ref{equ:fdmscaling}.  Thus, one can find numerical values for
$\sigma$, $\tau$ and $T_c$, or $p_c$, respectively, by conducting a
$\chi^2$ optimization procedure for the parameter set for which the
log of the scaled yield, $\langle n_Z \rangle / q_0 Z^{-\tau}$, as a
function of the scaled control parameter, $\epsilon Z^\sigma$,
collapses on a single straight line best fit.

The result of this optimization procedure for the unfiltered model
calculation is shown in Fig.~\ref{fig:scaled_perc}. The values of
the critical parameters extracted are $\sigma=0.5 \pm 0.1$, $\tau=2.18
\pm 0.01$, in good agreement with the accepted values of standard 3d 
percolation on infinite lattices, $\tau=2.18$ and $\sigma=0.45$.  We 
also find $p_c=0.65$.  This shows that $p_c$, unlike $\tau$ and 
$\sigma$, is strongly affected by 
finite size scaling corrections, in accordance with the findings of 
\cite{Bau84,Har00}.


Having shown that the method yields reasonable results in a known
case, we apply it to the determination of the critical
parameters of the ISiS data.  In previous analyses of this kind, no
corrections for sequential decays, feeding, population of particle
unstable resonances, and all other final state modifications of the
charge yield spectrum were considered (see \cite{Ell01}).  In
addition, all detector acceptance corrections were neglected.  We have
paid particular attention to these effects in the work presented here.

\begin{figure}
  \resizebox*{1.5\columnwidth}{!}{\includegraphics{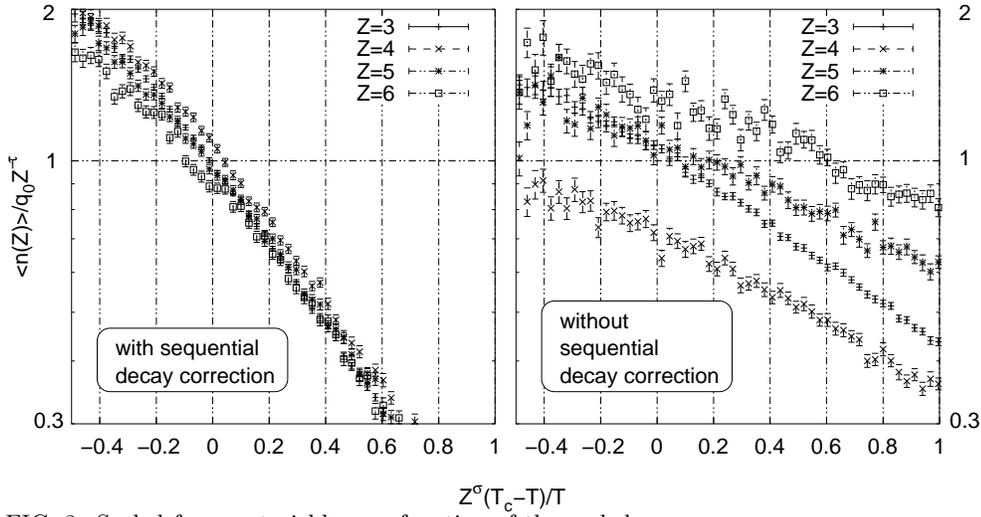}}
  \caption{Scaled fragment yields as a function of the scaled control
    parameter for $Z$ = 3, 4, 5, 6. The left hand side shows the
    results of the correct inclusion of secondary decay corrections,
    and the right hand side shows the best fit possible when omitting
    these corrections. }
  \label{fig:scaled_exp}
\end{figure}

\begin{figure}
  \resizebox*{0.8\columnwidth}{!}{\includegraphics{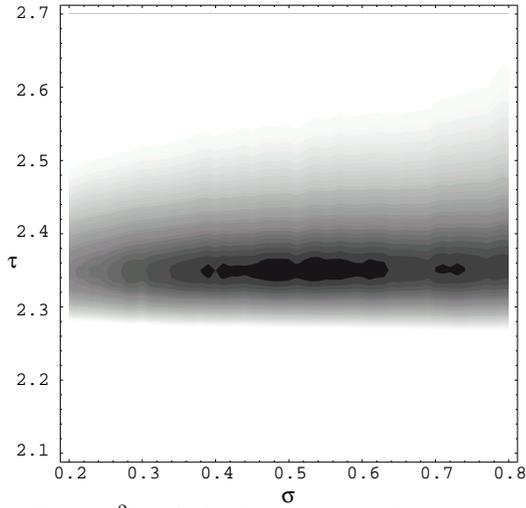}}
  \caption{$\chi^2$ optimization contours for the corrected ISiS
    data. A value of $T_c=8.3$ MeV was used.}
  \label{fig:fdmexpcont}
\end{figure}

To estimate the corrections for sequential decays, we start with our
model calculations presented in the previous sections.  These
calculations can reproduce almost all features of the data and in
particular the charge yield spectrum, after detector and final state
interaction corrections.  Since we know the model yields before and
after the corrections, we can extract the charge resolved correction
factors.  These factors are then applied to the experimental data. 
The result of the resulting $\chi^2$ optimization procedure is shown
in the left hand side of Figure \ref{fig:scaled_exp}.  The values of
the critical parameters obtained are $\sigma=0.5 \pm 0.1$, $\tau =
2.35 \pm 0.05$, and $T_c=8.3 \pm 0.2$ MeV. The contours of the
$(\sigma, \tau)-\chi^2$ fit are shown in Figure \ref{fig:fdmexpcont}
for $T_c=8.3$ MeV.

If one neglects the corrections for detector acceptance and sequential
decays, then there is no way that the yields for different light IMFs can
be collapsed onto a single scaling graph. On the right hand side of
Figure \ref{fig:scaled_exp} we show the best fit result of the
$\chi^2$ optimization for that case. It is obvious that the collapse
is not achieved.
This comparison can also be made for $Z>6$.  The 
ISiS data set has elementally resolved yields for $Z<17$.  But 
the effects of final state sequential decay corrections is strongest 
for the lightest element, and we thus restrict ourselves to show these 
here.

Summarizing, a three-step percolation model for nuclear
multi-fragmentation reactions has been introduced.  In order to reduce
unnecessary model dependences we have chosen to utilitize the
information on source size and excitation energy deposition provided
in the experiment.  For the fragmentation part of the model we use the
well-known percolation approach.  Particular attention is paid to the
effects of detector acceptance and sequential decays.  We find that
our calculations are in very good agreement with the data.  Since the
infinite size limit of the model contains a second order continuous
phase transition for a certain range of excitation energies that is
covered by a subset of the events in the present data set, we
interpret this agreement as strong circumstantial evidence for a
continuous phase transition in nuclear matter.  This interpretation is
supported by a scaling analysis.  We find that the data show very
strong scaling behavior, as expected in the vicinity of the critical
point.  The critical parameters extracted from a $\chi^2$ optimization
procedure have the values $\sigma=0.5 \pm 0.1$, $\tau=2.35 \pm 0.05$,
and $T_c=8.3 \pm 0.2$ MeV.

This work was supported by NSF grants INT-9981342 and PHY-0070818, the
US Department of Energy, and NSERC of Canada.  MKB also received
funding from the Studienstiftung des Deutschen Volkes.  WB
acknowledges support from an Alexander-von-Humboldt Foundation
Distinguished Senior U.S. Scientist Award.  The work reported on here
was part the Master's thesis of MKB.

\end{document}